\documentclass[11pt]{article}

\usepackage[margin=1in]{geometry}
\usepackage{amsmath}
\usepackage{array}
\usepackage{booktabs}
\usepackage{enumitem}
\usepackage{graphicx}
\usepackage{xcolor}
\usepackage{listings}
\usepackage{placeins}
\usepackage[numbers]{natbib}
\usepackage{hyperref}
\usepackage{microtype}

\definecolor{jsonstring}{RGB}{0,92,75}
\definecolor{jsonnumber}{RGB}{128,64,0}
\definecolor{jsonliteral}{RGB}{120,0,120}
\lstdefinelanguage{json}{
    morestring=[b]",
    stringstyle=\color{jsonstring},
    showstringspaces=false,
    literate=
     *{0}{{{\color{jsonnumber}0}}}{1}
      {1}{{{\color{jsonnumber}1}}}{1}
      {2}{{{\color{jsonnumber}2}}}{1}
      {3}{{{\color{jsonnumber}3}}}{1}
      {4}{{{\color{jsonnumber}4}}}{1}
      {5}{{{\color{jsonnumber}5}}}{1}
      {6}{{{\color{jsonnumber}6}}}{1}
      {7}{{{\color{jsonnumber}7}}}{1}
      {8}{{{\color{jsonnumber}8}}}{1}
      {9}{{{\color{jsonnumber}9}}}{1}
      {true}{{{\color{jsonliteral}true}}}{4}
      {false}{{{\color{jsonliteral}false}}}{5}
      {null}{{{\color{jsonliteral}null}}}{4}
}
\lstdefinestyle{boxlisting}{
    basicstyle=\footnotesize\ttfamily,
    breaklines=true,
    columns=fullflexible,
    frame=single,
    rulecolor=\color{black},
    framesep=5pt,
    aboveskip=0.7\baselineskip,
    belowskip=0.7\baselineskip,
    xleftmargin=0pt,
    xrightmargin=0pt
}
\lstdefinestyle{compactbox}{
    style=boxlisting,
    basicstyle=\scriptsize\ttfamily,
    framesep=4pt,
    aboveskip=0.6\baselineskip,
    belowskip=0.9\baselineskip
}

\title{SimpleWikiSearch: A Clean Offline Wikipedia Environment for Agentic Search}
\author{
Guanming Xiong \quad Penghui Zhang \\
Peking University \\
\texttt{gm\_xiong@pku.edu.cn} \quad \texttt{zhangph@pku.edu.cn}
}
\date{}

\begin{document}

\maketitle

\begin{abstract}
Large language model (LLM)-based agentic search systems are often evaluated as if the underlying LLM were the only component that matters, yet their measured performance also depends on the surrounding search environment: the Wikipedia snapshot, preprocessing pipeline, chunking policy, retrieval backend, tool schema, observation format, and answer submission rule. These details are frequently under-specified, making it difficult to compare results or reproduce reported baselines. We present SimpleWikiSearch, whose corpus construction, retrieval stack, tool contract, and evaluation protocol are explicit and runnable. The environment starts from a full English Wikipedia dump, cleans and chunks the corpus, builds keyword and dense retrieval indexes, and exposes a minimal tool interface consisting of \texttt{search}, \texttt{open\_url}, and \texttt{submit\_answer}. We report baseline results on six QA datasets using open-source LLMs and provide a random-300 subset for comparisons with closed-source commercial models. SimpleWikiSearch provides a domain-specific agent harness and a controlled offline environment for reproducible agentic-search evaluation. Its contribution is this specified reference setup, rather than a new agent algorithm. Code and data will be available at: \url{https://github.com/JimXiongGM/simple_wiki_search}.
\end{abstract}

\section{Introduction}

Open-domain and agentic question-answering systems are often compared based on final-answer scores, but the underlying search environment is rarely treated as a first-class experimental object. This is consequential because a system's behavior is shaped not only by the language model, but also by the knowledge snapshot, document processing pipeline, retrieval unit, index backend, tool schema and observation format.

A concrete example is the widely reused KILT/DPR-style Wikipedia setup.
KILT~\cite{Petroni-Fabio-NAACL-2021-KILT} grounds multiple knowledge-intensive tasks in a shared Wikipedia snapshot; its knowledge source is based on the 2019/08/01 English Wikipedia snapshot and contains 5.9M articles.
For passage retrieval, KILT adopts the DPR~\cite{Karpukhin-Vladimir-EMNLP-2020-DPR} passage-construction convention while using its own snapshot: DPR uses the December 20, 2018 English Wikipedia dump, removes semi-structured content such as tables, infoboxes, lists, and disambiguation pages, and splits articles into disjoint 100-word passages.
The resulting DPR corpus contains 21,015,324 passages; applying the same 100-word split to the KILT snapshot yields 22,220,793 passages.
These design choices were reasonable for passage-level retrievers and short-context readers, but they are not neutral defaults for modern large language model (LLM) based tool-using agents: both the retrieval-unit convention and the surrounding environment details need to be stated explicitly.

Prior work has already shown that short passage units can hide document-level information. \citet{Kamalloo-Ehsan-SIGIR-2023-DocReasoning} find that common open-domain QA benchmarks are heavily biased toward passage-level evidence and identify questions that require document-level reasoning rather than passage-only retrieval. From a modern RAG perspective, \citet{Jiang-Ziyan-arXiv-2024-LongRAG} further argue that 100-word retrieval units force the retriever to find a small ``needle'' in a large corpus, while short chunks can become semantically incomplete because of document truncation and information loss.
Although we do not adopt their long-context RAG design, their analysis supports the same conclusion: corpus construction and retrieval-unit granularity should be reported and revisited as model context windows and agent interfaces change.

The same under-specification appears in agentic-search papers.
Many systems state that an agent can search Wikipedia, but leave the environment under-specified: which dump or derived database is used, how pages are cleaned and segmented, how search results are formatted, how observations are truncated, and whether multiple tools can be called in a single step.
These details can change both retrieval behavior and reasoning behavior, so an agentic-search baseline should specify them explicitly.

We present SimpleWikiSearch as a clean baseline for agentic search over offline Wikipedia. It specifies the corpus snapshot, page cleaning, section-aware chunking, keyword and dense indexes, tool schema, observation format, and evaluation protocol, and releases them with the paper.

Our contributions are as follows:
\begin{itemize}[nolistsep]
    \item We specify an end-to-end offline Wikipedia environment for agentic QA, including corpus layout, retrieval services, tool definitions, and the agent interaction protocol.
    \item We report full-test and random-300 benchmark results on six QA datasets using open-source LLMs, with closed-source commercial-model results on the random-300 subset.
    \item We provide reproducible analysis scripts for result tables and tool-use statistics, so that reporting is tied to executable code rather than hand-maintained numbers.
    \item We release all inference result files, including saved message trajectories, tool observations, predictions, and evaluation fields, to facilitate community analysis.
\end{itemize}

\section{Environment}

\subsection{Corpus and Retrieval}

The current release uses the official English Wikipedia dump \texttt{enwiki-20260601}. The preprocessing pipeline converts wikitext to a cleaned plain-text representation (lightweight Markdown: section headings and tables when present, with explicit link markup removed) before indexing. The released retrieval indexes cover 7,189,602 Wikipedia articles.

Chunks are defined by token length after tokenization, rather than by a fixed word count. We use the \texttt{Qwen3-Embedding-0.6B}~\cite{Zhang-Yanzhao-arXiv-2025-Qwen3Embedding} tokenizer and merge sections toward a target length of 1536 tokens, with a 20\% tolerance band and sentence-boundary preservation. This produces 10,837,506 chunks for dense embedding and retrieval. The design differs from legacy 100-word KILT/DPR-style passages: the agent observes longer snippets that preserve page titles, section structure, and table-like content when available. Each chunk stores a Wikipedia page id, page title, chunk id (of the form \texttt{<pageid>\_<chunk\_index>}), and a zero-based internal chunk index; search results expose offline-stable URLs of the form \texttt{/wiki/<pageid>\#chunk-N}, where \texttt{N} is one-based.

Table~\ref{tab:corpus-construction} summarizes the released corpus and index configuration. The retrieval stack contains a Tantivy\footnote{\url{https://github.com/quickwit-oss/tantivy}} keyword index over chunk text and titles, plus a FAISS HNSW dense index built from \texttt{Qwen3-Embedding-0.6B} embeddings. Dense search returns chunk ids only; the corresponding text is resolved by looking up those ids in the Tantivy chunk store. The two indexes therefore share a single chunk-id space over the same 7,189,602 articles. The default search method is reciprocal rank fusion (RRF) over keyword and vector retrieval, while keyword-only and vector-only modes remain exposed for controlled comparisons.

\begin{table}[!ht]
\centering
\small
\setlength{\tabcolsep}{4pt}
\renewcommand{\arraystretch}{0.95}
\begin{tabular}{p{0.24\linewidth}p{0.68\linewidth}}
    \toprule
    \textbf{Component} & \textbf{Released setting} \\ \midrule
    Wikipedia snapshot & Official English Wikipedia dump \texttt{enwiki-20260601}; released indexes cover 7,189,602 articles and 10,837,506 chunks. \\
    Text representation & Wikitext is cleaned before indexing; search and open return plain text with section structure, not raw HTML or short KILT/DPR-style passages. \\
    Tokenizer and chunking & \texttt{Qwen3-Embedding-0.6B} tokenizer; section-aware \texttt{chunk-1536} construction with 20\% tolerance and sentence-boundary preservation. \\
    Chunk metadata & Page id, page title, chunk id (\texttt{<pageid>\_<chunk\_index>}), zero-based internal chunk index, total chunks, and text are retained; exposed URLs use one-based \texttt{\#chunk-N}. \\
    Keyword index & Tantivy indexes chunk text and document titles, with a separate title index for page lookup and URL resolution; the released keyword and dense indexes share the same chunk ids. \\
    Dense embeddings & \texttt{Qwen3-Embedding-0.6B} embeddings are computed for all chunks, producing 1024-dimensional vectors. \\
    Dense index & FAISS HNSW stores one fp16 vector per chunk (dim~$=1024$), with a side table mapping vector row ids to chunk ids. \\
    Text resolution & Vector hits return chunk ids only; full chunk text and titles are resolved from Tantivy using those ids. \\
    Default retrieval & RRF fusion over keyword and dense candidates; \texttt{keywords} and \texttt{vector} modes remain selectable for ablations. \\ \bottomrule
\end{tabular}
\caption{Wikipedia corpus construction and retrieval-index configuration used in this release.}
\label{tab:corpus-construction}
\end{table}

\subsection{Tool Interface}

The QA agent is given exactly three OpenAI-style function tools. The tool contract is intentionally minimal: \texttt{search} retrieves ranked snippets, \texttt{open\_url} opens a returned Wikipedia URL, and \texttt{submit\_answer} terminates the episode. Returned URLs use the \texttt{en.wikipedia.org} host for readability, but they resolve against the offline index rather than the live web.

Search observations are ranked hits. Each hit includes a title, an openable offline Wikipedia chunk URL, and a snippet:

\begin{center}
\begin{minipage}{0.9\linewidth}
\begin{lstlisting}[style=compactbox]
1. [Brassia]
   (https://en.wikipedia.org/wiki/1171017#chunk-1)
   - Section 0: Preamble Brassia is a genus of orchids classified in the subtribe Oncidiinae. It is native to Mexico, Central America, the West Indies, and northern South America...

2. [Brassia allenii]
   (https://en.wikipedia.org/wiki/49420403#chunk-1)
   - Section 0: Preamble
     Brassia allenii is a species of orchid. It is native to Honduras and Panama.
\end{lstlisting}
\end{minipage}
\end{center}

Opening a chunk URL returns that chunk with title and position. Opening a page URL without \texttt{\#chunk-N} returns the full article text.

\begin{center}
\begin{minipage}{0.9\linewidth}
\begin{lstlisting}[style=compactbox]
# Scout Tufankjian (chunk 1/2)

Section 0: Preamble
Scout Tufankjian is an Armenian-American photojournalist and author based in Brooklyn, New York. She is known for her photos of American president Barack Obama during his first presidential campaign...

Section 1: Early life and education
Tufankjian was born in 1977 in Boston, Massachusetts...
\end{lstlisting}
\end{minipage}
\end{center}

The complete JSON schema passed to the model is:

\begin{lstlisting}[style=boxlisting,language=json]
[
  {
    "type": "function",
    "function": {
      "name": "search",
      "description": "Search offline knowledge base and return ranked hits with URLs and one-line snippets. method=\"rrf\" gives hybrid recall/balance (default), method=\"keywords\" is lexical exact-match friendly, and method=\"vector\" is semantic similarity friendly.",
      "parameters": {
        "type": "object",
        "properties": {
          "query": {"type": "string", "description": "Search query."},
          "top_k": {"type": "integer", "description": "Number of hits to return.", "default": 10, "minimum": 1},
          "only_title": {"type": "boolean", "description": "Whether to only search article titles.", "default": false},
          "method": {"type": "string", "description": "Search method: \"rrf\" (hybrid/balanced, recommended default), \"keywords\" (lexical exact-token matching), \"vector\" (semantic embedding similarity).", "enum": ["rrf", "keywords", "vector"], "default": "rrf"}
        },
        "required": ["query"]
      }
    }
  },
  {
    "type": "function",
    "function": {
      "name": "open_url",
      "description": "Open a URL. If URL has #chunk-N suffix, returns that chunk; remove the #chunk-N suffix to fetch the full article.",
      "parameters": {
        "type": "object",
        "properties": {"url": {"type": "string", "description": "URL to open."}},
        "required": ["url"]
      }
    }
  },
  {
    "type": "function",
    "function": {
      "name": "submit_answer",
      "description": "Submit the final answer and end the episode: a short exact string, or one Markdown table when the question requires it.",
      "parameters": {
        "type": "object",
        "properties": {"answer": {"type": "string", "description": "Final answer only (short string or one Markdown table), no explanation."}},
        "required": ["answer"]
      }
    }
  }
]
\end{lstlisting}

\subsection{Agent Protocol}

The baseline agent uses an OpenAI-compatible chat-completion server. At each round, the model receives the conversation history and the tool schema, may request tool calls automatically, and may call multiple independent tools in one step. The runner caps each episode at 20 model rounds and truncates any assistant step to at most five tool calls. If the model emits no tool call, the runner adds a reminder to continue with tools or finish with \texttt{submit\_answer}. These choices are intentionally simple; their main purpose is to make the environment contract explicit.

\section{Benchmark Setup}

\begin{table}[ht]
\centering
\scalebox{1}{
\begin{tabular}{llccc}
    \toprule
    \textbf{Dataset} & \textbf{Split} & \textbf{$n$} & \textbf{Avg. Q words} & \textbf{Alias \%} \\ \midrule
    2Wiki & dev & 12,576 & 12.0 & 84.3 \\
    HotpotQA & dev-fullwiki & 7,405 & 15.7 & 0.0 \\
    MuSiQue & dev & 2,417 & 18.1 & 28.1 \\
    FRAMES & all & 824 & 27.6 & 0.0 \\
    PopQA & all & 14,267 & 6.6 & 100.0 \\
    Bamboogle & all & 125 & 11.9 & 0.8 \\ \bottomrule
\end{tabular}
}
\caption{Statistics of the datasets used in our experiments. Alias coverage is the percentage of examples that provide at least one answer alias.}
\label{tab:dataset-stats}
\end{table}

We evaluate on 2WikiMultiHopQA~\cite{Ho-Xanh-COLING-2020-2WikiMultiHopQA}, HotpotQA~\cite{Yang-Zhilin-EMNLP-2018-HotpotQA}, MuSiQue~\cite{Trivedi-Harsh-TACL-2022-MuSiQue}, FRAMES~\cite{Krishna-Satyapriya-arXiv-2024-FRAMES}, PopQA~\cite{Mallen-Alex-ACL-2023-PopQA}, and Bamboogle~\cite{Press-Ofir-EMNLP-2023-Bamboogle}. Table~\ref{tab:dataset-stats} reports the processed evaluation splits used by the runner. The full-test setting evaluates the available gold-answer split for each dataset. The random-300 setting evaluates 300 examples per dataset, except Bamboogle, which contains only 125 examples.

The open-source LLM baselines are \texttt{Qwen3.5-4B} and \texttt{Qwen3.5-9B}~\cite{Qwen-Team-2026-Qwen3.5}. The random-300 subset additionally includes closed-source commercial models \texttt{deepseek-v4-pro} and \texttt{gpt-5.4-2026-03-05}. Unless noted otherwise, agents use the default RRF search method, temperature $0.7$, top-$p$ $0.95$, a 20-round episode cap, and at most five tool calls per assistant step. Open-source LLM inference is served locally with SGLang~\cite{Zheng-Lianmin-NeurIPS-2024-SGLang} on a single NVIDIA A100 80GB GPU under CUDA~13.0; the reported runtimes therefore reflect this hardware and serving stack.

We report two complementary metrics as percentages. Token-level F1 follows the standard open-domain QA normalization (lowercasing, punctuation and article removal) and takes the maximum over the gold answer and any provided aliases. LLM-judged accuracy is computed using \texttt{gpt-5.4-mini-2026-03-17} with a fixed binary prompt that credits factually equivalent answers, including paraphrases and answers embedded in longer responses. The two metrics can diverge when a correct answer is verbose or lexically distant from the gold string; we therefore report both rather than collapsing evaluation to a single score. The judge system prompt and user prompt template are:

\begin{center}
\begin{minipage}{0.9\linewidth}
\begin{lstlisting}[style=compactbox]
System:
You are an expert QA evaluator. Judge factual correctness only; ignore style, grammar, and punctuation.

User:
Decide whether the prediction is factually correct.
- Credit answers embedded in a longer response.
- Treat equivalent dates, spellings, and name variants as correct.
- Mark incorrect if it contradicts the golden answer or fails to answer.

Question: <question>
Golden answer (any line below is acceptable):
<main answer>
- <alias_1>
- ...
Prediction: <prediction>

Give one short sentence of reasoning, then write exactly True or False on the last line.
\end{lstlisting}
\end{minipage}
\end{center}

\section{Results}

Table~\ref{tab:score-full} reports full-test results. To keep the table readable, datasets are shown in two panels separated by a double horizontal rule while remaining within a single table. Under the LLM-judged metric, the open-source 9B model improves over the 4B model on most datasets, with the largest gains on MuSiQue, FRAMES, and Bamboogle; HotpotQA is an exception, where 9B is slightly lower. The comparison also shows that final-answer F1 and judge accuracy do not always align, which is expected for datasets where correct answers are often paraphrased or embedded in longer strings.

\begin{table*}[htp]
\centering
\scalebox{1}{
\begin{tabular}{lrrrrrrrrr}
    \toprule
    \textbf{Method} & \multicolumn{3}{c}{\textbf{2Wiki}} & \multicolumn{3}{c}{\textbf{HotpotQA}} & \multicolumn{3}{c}{\textbf{MuSiQue}} \\
    \cmidrule(lr){2-4} \cmidrule(lr){5-7} \cmidrule(lr){8-10}
     & \textbf{$n$} & \textbf{F1} & \textbf{Judge} & \textbf{$n$} & \textbf{F1} & \textbf{Judge} & \textbf{$n$} & \textbf{F1} & \textbf{Judge} \\ \midrule
    Qwen3.5-4B & 12,576 & 76.44 & 86.28 & 7,405 & 53.95 & 70.93 & 2,417 & 29.07 & 35.37 \\
    Qwen3.5-9B & 12,576 & 75.59 & 87.75 & 7,405 & 55.18 & 69.18 & 2,417 & 32.58 & 40.50 \\
    \midrule \midrule
    \textbf{Method} & \multicolumn{3}{c}{\textbf{FRAMES}} & \multicolumn{3}{c}{\textbf{PopQA}} & \multicolumn{3}{c}{\textbf{Bamboogle}} \\
    \cmidrule(lr){2-4} \cmidrule(lr){5-7} \cmidrule(lr){8-10}
     & \textbf{$n$} & \textbf{F1} & \textbf{Judge} & \textbf{$n$} & \textbf{F1} & \textbf{Judge} & \textbf{$n$} & \textbf{F1} & \textbf{Judge} \\ \midrule
    Qwen3.5-4B & 824 & 42.35 & 54.85 & 14,267 & 53.93 & 65.56 & 125 & 66.24 & 73.60 \\
    Qwen3.5-9B & 824 & 48.12 & 60.80 & 14,267 & 56.57 & 65.70 & 125 & 69.30 & 73.60 \\ \bottomrule
\end{tabular}
}
\caption{Full-test benchmark results.}
\label{tab:score-full}
\end{table*}

Table~\ref{tab:score-sample} reports the random-300 subset using the same two-panel layout. This subset is intended for lower-cost comparisons with closed-source commercial models. The commercial models improve judge accuracy on several datasets, especially FRAMES and Bamboogle, while the open-source LLMs remain competitive on entity-centric and simpler multi-hop Wikipedia QA. The same table also shows a clear F1--judge discrepancy: on 2Wiki, for example, the commercial models achieve lower lexical F1 but higher judge accuracy than the open-source LLMs, consistent with longer or paraphrased answers that a lexical metric under-credits.

\begin{table*}[htp]
\centering
\scalebox{1}{
\begin{tabular}{lrrrrrrrrr}
    \toprule
    \textbf{Method} & \multicolumn{3}{c}{\textbf{2Wiki}} & \multicolumn{3}{c}{\textbf{HotpotQA}} & \multicolumn{3}{c}{\textbf{MuSiQue}} \\
    \cmidrule(lr){2-4} \cmidrule(lr){5-7} \cmidrule(lr){8-10}
     & \textbf{$n$} & \textbf{F1} & \textbf{Judge} & \textbf{$n$} & \textbf{F1} & \textbf{Judge} & \textbf{$n$} & \textbf{F1} & \textbf{Judge} \\ \midrule
    Qwen3.5-4B & 300 & 77.73 & 86.33 & 300 & 56.99 & 71.33 & 300 & 28.89 & 33.67 \\
    Qwen3.5-9B & 300 & 82.29 & 89.67 & 300 & 59.46 & 73.00 & 300 & 33.09 & 43.67 \\
    deepseek-v4-pro & 300 & 67.30 & 90.67 & 300 & 60.79 & 79.00 & 300 & 31.44 & 44.67 \\
    gpt-5.4-2026-03-05 & 300 & 67.88 & 90.33 & 300 & 63.37 & 87.67 & 300 & 39.94 & 58.33 \\
    \midrule \midrule
    \textbf{Method} & \multicolumn{3}{c}{\textbf{FRAMES}} & \multicolumn{3}{c}{\textbf{PopQA}} & \multicolumn{3}{c}{\textbf{Bamboogle}} \\
    \cmidrule(lr){2-4} \cmidrule(lr){5-7} \cmidrule(lr){8-10}
     & \textbf{$n$} & \textbf{F1} & \textbf{Judge} & \textbf{$n$} & \textbf{F1} & \textbf{Judge} & \textbf{$n$} & \textbf{F1} & \textbf{Judge} \\ \midrule
    Qwen3.5-4B & 300 & 42.81 & 55.67 & 300 & 53.32 & 65.00 & 125 & 66.24 & 73.60 \\
    Qwen3.5-9B & 300 & 46.20 & 57.67 & 300 & 57.95 & 64.00 & 125 & 69.30 & 73.60 \\
    deepseek-v4-pro & 300 & 67.04 & 85.67 & 300 & 53.05 & 73.33 & 125 & 84.47 & 91.20 \\
    gpt-5.4-2026-03-05 & 300 & 65.63 & 84.33 & 300 & 56.60 & 73.00 & 125 & 84.01 & 92.80 \\ \bottomrule
\end{tabular}
}
\caption{Random-300 benchmark results.}
\label{tab:score-sample}
\end{table*}

\section{Analysis}

Final-answer scores hide substantial differences in how agents use the environment. Tables~\ref{tab:tool-stats-qwen354b} and~\ref{tab:tool-stats-qwen359b} summarize the full-test interaction traces for the two open-source LLMs. We report the average number of model rounds, search calls, open calls, successful submit rate, total tokens (prompt plus completion), and wall-clock runtime per question.

\begin{table}[htp]
\centering
\scalebox{1}{
\begin{tabular}{lrrrrrrr}
    \toprule
    \textbf{Dataset} & \textbf{$n$} & \textbf{Rounds} & \textbf{Search} & \textbf{Open} & \textbf{\begin{tabular}[c]{@{}c@{}}Submit\\\%\end{tabular}} & \textbf{\begin{tabular}[c]{@{}c@{}}Total\\tokens\end{tabular}} & \textbf{Sec.} \\ \midrule
    2Wiki & 12,576 & 4.49 & 3.49 & 0.48 & 96.13 & 24,070 & 11.8 \\
    HotpotQA & 7,405 & 7.17 & 5.53 & 0.87 & 85.52 & 55,399 & 21.6 \\
    MuSiQue & 2,417 & 11.42 & 9.56 & 1.20 & 68.47 & 103,362 & 37.0 \\
    FRAMES & 824 & 11.68 & 9.46 & 1.50 & 66.26 & 114,771 & 39.8 \\
    PopQA & 14,267 & 4.58 & 2.71 & 0.56 & 96.29 & 24,491 & 11.6 \\
    Bamboogle & 125 & 7.64 & 5.36 & 1.01 & 88.80 & 54,395 & 21.7 \\ \bottomrule
\end{tabular}
}
\caption{Tool-use statistics on full-test runs for Qwen3.5-4B. Except for $n$, values are averages per question; total tokens count prompt and completion tokens.}
\label{tab:tool-stats-qwen354b}
\end{table}

\begin{table}[ht]
\centering
\scalebox{1}{
\begin{tabular}{lrrrrrrr}
    \toprule
    \textbf{Dataset} & \textbf{$n$} & \textbf{Rounds} & \textbf{Search} & \textbf{Open} & \textbf{\begin{tabular}[c]{@{}c@{}}Submit\\\%\end{tabular}} & \textbf{\begin{tabular}[c]{@{}c@{}}Total\\tokens\end{tabular}} & \textbf{Sec.} \\ \midrule
    2Wiki & 12,576 & 4.17 & 3.38 & 0.46 & 96.79 & 24,728 & 15.3 \\
    HotpotQA & 7,405 & 6.82 & 5.24 & 0.82 & 87.83 & 56,234 & 29.4 \\
    MuSiQue & 2,417 & 10.87 & 8.76 & 1.07 & 73.31 & 104,643 & 54.0 \\
    FRAMES & 824 & 11.29 & 8.89 & 1.43 & 71.00 & 119,330 & 65.6 \\
    PopQA & 14,267 & 4.20 & 2.50 & 0.64 & 97.43 & 23,990 & 14.8 \\
    Bamboogle & 125 & 7.55 & 5.58 & 0.86 & 90.40 & 58,021 & 41.7 \\ \bottomrule
\end{tabular}
}
\caption{Tool-use statistics on full-test runs for Qwen3.5-9B. Except for $n$, values are averages per question; total tokens count prompt and completion tokens.}
\label{tab:tool-stats-qwen359b}
\end{table}

These statistics are part of the benchmark contract. Harder multi-hop datasets such as MuSiQue and FRAMES require roughly twice as many rounds and search calls as 2Wiki or PopQA, and their successful submission rates drop into the 66--73\% range, indicating that many episodes end by hitting the round budget rather than by a successful answer submission. Future systems may improve final QA accuracy through stronger reasoning, stronger retrieval, more effective tool use, or a larger interaction budget. Reporting interaction statistics makes those tradeoffs explicit.

\section{Conclusion}

SimpleWikiSearch provides a clean, specified baseline for agentic search over offline Wikipedia. Its primary contribution is a reproducible environment contract covering corpus construction, retrieval backend, tool interface, agent loop, result schema, evaluation, and interaction statistics, which makes agentic search results easier to reproduce and compare.

\bibliographystyle{plainnat}
\bibliography{references}

\end{document}